\documentstyle[preprint,pre,aps]{revtex}
\begin{document}
\draft 
\title{%
Two-dimensional cellular automaton model of
traffic flow with open boundaries}
\author{Shin-ichi Tadaki\cite{email}}
\address{Department of Information Science,
Saga University, Saga 840, Japan}
\date{\today}
\maketitle
\begin{abstract}
A two-dimensional cellular automaton model of traffic flow
with open boundaries are investigated by computer
simulations.  The outflow of cars from the system and the
average velocity are investigated.  The time sequences of
the outflow and average velocity have flicker noises in a
jamming phase.  The low density behavior are discussed with
simple jam-free approximation.
\end{abstract}
\narrowtext
\section{Introduction}

Models of traffic flow have relations to wide varieties of
physical systems. A traffic flow system is one of asymmetric
exclusion processes.  They are non-trivial statistical
mechanical systems because of lack of detailed balance.
Studies of these non-trivial system explore the profound
structure of statistical mechanics.  Studies of exclusively
interacting particle systems like traffic flow also relates
to equilibrium and non-equilibrium properties of granular
flows, surface growth, dynamics of defects in solids, and so
on. The model I will discuss here may be one of the simplest
examples of non-equilibrium colliding granular flows.

Traffic flow problems have been studied mainly through fluid
dynamics, car-following models, coupled map lattice models,
and cellular automaton (CA) models.  Many attempts have been
made to apply CA modeling to complex phenomena including
fluid because of computational simplicity.  Cellular
automaton modeling of traffic flow is one of the recently
developing area.  One of the simplest CA models of traffic
flow in one-way expressway is the rule-184 elementary
CA\cite{Wolfram}, which is a simple asymmetric exclusion
rule.  In spite of the simplicity of the model, it shows a
phase transition from a freely moving phase at low vehicle
density to a jamming phase at high vehicle density.  The
computational simplicity of CA models also enables us to
take many realistic features of traffic problems into
account.  More realistic models considering speed variation
of cars or effects of blockades have been investigated in
one-dimensional models\cite{Nagel:1,Nagel:2,SS,Yukawa}.
$1/f$ fluctuation has been observed in both actual
expressways \cite{Musha} and models\cite{Nagel:2,T2}.  Self
organized criticality has also been studied\cite{Nagel:3}.

Traffic networks, for example a traffic system of a whole
city or an expressway network, consist of many complicated
ingredients.  It is very hard to model the whole features of
traffic networks.  Two-dimensional CA models of traffic
flow, therefore, are very abstract models of traffic
networks.  One of the simplest two-dimensional CA model of
traffic flow has been investigated by Biham, Middleton and
Levine (BML)\cite{BML}.  Their model is a simple extension
of the rule-184 CA to two dimension.  Cars are distributed
on a square lattice of $N\times N$ sites with periodic
boundaries with both in the horizontal and vertical
directions.  They found a sharp transition between a freely
moving phase at low vehicle density and a jamming phase at
high vehicle density.  The characteristics of the transition
were studied by Nagatani\cite{Nagatani:1} and Fukui and
Ishibashi\cite{Fukui,Ishibashi}.  Two type of jam phase was
discussed by Tadaki and Kikuchi\cite{Tadaki:1,Tadaki:2}.
The model has been extended to take into account the
probability of changes in vehicle
directions\cite{Cuest,Molera}.

Two-dimensional CA models of traffic flow show many
physically interesting phenomena, phase transitions and self
organization. Cellular automaton modeling of traffic
systems, however, is a toy model. It should be clarified
what features strongly depend on the model itself.
Characteristic features of the BML model are, for example,
deterministic dynamics, periodic boundaries, restrictions on
the car destination, road arrangement without traffic
queue\cite{Freund} and so on. In this paper, open boundary
conditions instead of periodic ones are used to investigate
the emergence of traffic jam in two-dimensional CA model.

The organization of this paper is as follows: The model is
given in \ref{Model}.  The dynamics is described with binary
arrays. The outflow of cars from the system is investigated
in \ref{Outflow}.  The jam-free approximation is discussed.
In \ref{Velocity} the average velocity of cars is
investigated.  Section \ref{Discussion} is devoted for
discussions.

\section{Model}\label{Model}
The model is the same as the model-I of BML except the
boundary conditions.  On the contrary to the original BML
model, cars are injected probabilistically both the left and
lower boundaries of the system and flow out
deterministically from both the right and upper boundaries.

Up-directed and right-directed cars are exclusively
distributed in $N\times N$ square lattice.  Each site is
empty or occupied by one up-directed or right-directed car.
Cars can move one step at a time if and only if the adjacent
site in the destination is empty.  There is a traffic light
controlling the whole system as up-directed cars can move
only at even time steps and right-directed cars can only at
odd time steps.

The number of right-directed (up-directed) cars at a time
$t$ and a position $\vec{r}=(i,j)$ ($1\le i,j\le N$) is
expressed by a binary array $\mu_{\vec{r}}(t)=\lbrace
0,1\rbrace$ ($\nu_{\vec{r}}(t)=\lbrace
0,1\rbrace$)\cite{Molera}.  The bulk dynamics, namely
dynamics for bulk sites $(i,j)$ ($1\le i\le N$, $1<j<N$ for
up-directed and $1<i<N$, $1\le j\le N$ for right-directed
cars) can be expressed as
\begin{eqnarray}
\mu_{\vec{r}}(t+1)&=&
\sigma(t)\mu_{\vec{r}}(t)
\left\lbrace\mu_{\vec{r}+\vec{x}}(t)
             +\nu_{\vec{r}+\vec{x}}(t)\right\rbrace
\nonumber\\
&&+\sigma(t)\left\lbrace1-\mu_{\vec{r}}(t)\right\rbrace 
 \left\lbrace1-\nu_{\vec{r}}(t)\right\rbrace 
\mu_{\vec{r}-\vec{x}}(t)\nonumber\\
&&+\left\lbrace1-\sigma(t)\right\rbrace \mu_{\vec{r}}(t),
\label{right-bulk}\\
\nu_{\vec{r}}(t+1)&=&
\left\lbrace1-\sigma(t)\right\rbrace \nu_{\vec{r}}(t)
\left\lbrace\mu_{\vec{r}+\vec{y}}(t)
           +\nu_{\vec{r}+\vec{y}}(t)\right\rbrace
\nonumber\\
&&+\left\lbrace1-\sigma(t)\right\rbrace 
\left\lbrace1-\mu_{\vec{r}}(t)\right\rbrace 
\left\lbrace1-\nu_{\vec{r}}(t)\right\rbrace 
\mu_{\vec{r}-\vec{y}}(t)\nonumber\\
&&+\sigma(t)\nu_{\vec{r}}(t),\label{up-bulk}
\end{eqnarray}
where $\vec{x}$ and $\vec{y}$ denote unit vectors of right
and up directions respectively.  The binary function
$\sigma(t)=t\bmod2$ represents the control by the traffic
light.  The condition $\mu_{\vec{r}}(t)\nu_{\vec{r}}(t)=0$
holds because one site can not be occupied both with up and
right cars simultaneously.

The first term \(\sigma(t)\mu_{\vec{r}}(t)
\left\lbrace\mu_{\vec{r}+\vec{x}}(t)
+\nu_{\vec{r}+\vec{x}}(t)\right\rbrace\) in
eq.~(\ref{right-bulk}) denotes that a right-directed car
remains at the site $\vec{r}$ if the right adjacent site is
occupied by a right-directed or up-directed car. The
injection of a right-directed car from the left adjacent
site is given by the second term
\(\sigma(t)\left\lbrace1-\mu_{\vec{r}}(t)\right\rbrace
\left\lbrace1-\nu_{\vec{r}}(t)\right\rbrace
\mu_{\vec{r}-\vec{x}}(t)\).  The last term
\(\left\lbrace1-\sigma(t)\right\rbrace \mu_{\vec{r}}(t)\)
shows that a right-directed car does not move at odd time
steps. The same decomposition of eq.~(\ref{up-bulk}) can be
done for the dynamics of up-directed cars.

Cars are injected from lower and left sides of the system.
If the site on the edges of the system is empty, a car is
injected with a probability $p$.  The injection of
right-directed cars on the left edge $\vec{r}=(1,j)$ ($1\le
j\le N$) is given by replacing the injection term (second
term) in eq.~(\ref{right-bulk}) with probabilistic
injection:
\begin{eqnarray}
\mu_{\vec{r}}(t+1)&=&
\sigma(t)\mu_{\vec{r}}(t)
\left\lbrace\mu_{\vec{r}+\vec{x}}(t)
                 +\nu_{\vec{r}+\vec{x}}(t)\right\rbrace
\nonumber\\
&&+\sigma(t)\left\lbrace1-\mu_{\vec{r}}(t)\right\rbrace 
\left\lbrace1-\nu_{\vec{r}}(t)\right\rbrace f(p)\nonumber\\
&&+\left\lbrace1-\sigma(t)\right\rbrace \mu_{\vec{r}}(t),
\label{right-in}
\end{eqnarray}
where $f(p)=\lbrace 0,1\rbrace$ is a function which returns
unity with a probability $p$.  The injection of up-directed
cars on the lower edge $\vec{r}=(i,1)$ ($1\le i\le N$) is
given by
\begin{eqnarray}
\nu_{\vec{r}}(t+1)&=&
\left\lbrace1-\sigma(t)\right\rbrace \nu_{\vec{r}}(t)
\left\lbrace\mu_{\vec{r}+\vec{y}}(t)
                 +\nu_{\vec{r}+\vec{y}}(t)\right\rbrace
\nonumber\\
&&+\left\lbrace1-\sigma(t)\right\rbrace 
\left\lbrace1-\mu_{\vec{r}}(t)
\right\rbrace \left\lbrace1-\nu_{\vec{r}}(t)\right\rbrace f(p)
\nonumber\\
&&+\sigma(t)\nu_{\vec{r}}(t).
\label{up-in}
\end{eqnarray}

Cars flow out from the upper and right edges of the system
deterministically.  The dynamical equations for cars on the
upper and rights edge are given by deleting the first terms
in eqs.~(\ref{right-bulk}) and (\ref{up-bulk}).  For sites
on the right edge $\vec{r}=(N,j)$ ($1\le j\le N$)
\begin{eqnarray}
\mu_{\vec{r}}(t+1)&=&
\sigma(t)\left\lbrace1-\mu_{\vec{r}}(t)\right\rbrace 
\left\lbrace1-\nu_{\vec{r}}(t)\right\rbrace 
\mu_{\vec{r}-\vec{x}}(t) 
\nonumber\\
&&+\left\lbrace1-\sigma(t)\right\rbrace \mu_{\vec{r}}(t)
\label{right-out}
\end{eqnarray}
gives the dynamics of right-directed cars on the right edge.
For sites on the upper edge $\vec{r}=(i,N)$ ($1\le i\le N$)
\begin{eqnarray}
\nu_{\vec{r}}(t+1)&=&
\left\lbrace1-\sigma(t)\right\rbrace 
\left\lbrace1-\mu_{\vec{r}}(t)\right\rbrace 
\left\lbrace1-\nu_{\vec{r}}(t)\right\rbrace 
\nu_{\vec{r}-\vec{y}}(t)
\nonumber\\
&&+\sigma(t)\nu_{\vec{r}}(t)
\label{up-out}
\end{eqnarray}
gives the dynamics of up-directed cars on the upper edge.

In the current simulations, the system has no car at the
initial time $t=0$.  Cars are injected with
eqs.~(\ref{right-in}) and (\ref{up-in}) probabilistically
and run deterministically obeying eqs.~(\ref{right-bulk})
and (\ref{up-bulk}).  If cars reach the edges of the system,
they flow out by eqs.~(\ref{right-out}) and (\ref{up-out}).
In the early traffic light cycles to $O(N)$, the front lines
of right-directed and up-directed cars collide to form a
global traffic jam configuration in case $p>p_c$ ($p_c$ is
discussed later).  The global jam is sorted out with the
maximum throughput, where the number of cars per site is
$\rho=2/3$.  Then new small jam clusters are created and
sorted out again and again.  Figure \ref{snap} shows a
snapshot of the system.  In the simulation, the system runs
$200N$ times ($100N$ traffic light cycles) from the initial
condition for relaxation.  And quantities discussed later
are observed for $200N<t\le400N$.

\section{Outflow}\label{Outflow}
The first quantity we observe is the outflow of cars from
the system.  By virtue of the dynamics,
eqs.~(\ref{right-out}) and (\ref{up-out}), the outflow is
the number of cars appearing on the upper and right edges of
the system.  The average outflow $\bar{p}_{\rm out}$ is
defined as the average number of cars appearing on the edges
per site and traffic light cycle.  The results of the
simulation is given in Fig.~\ref{AverageOutflow}.

In the low injection $p$ region, the system can be assumed
to be free from traffic jam.  In this case the injection
process will be controlled only by the number of cars which
stay on the lower and left edges of the system.  The number
of up-directed (right-directed) cars in each column (row) is
given by $\bar{p}_{\rm out}N$.  There are $2\bar{p}_{\rm
out}N$ cars on the left edge of the system.  These cars
prevent the car injection from the left side of the system,
and the remaining $(1-2\bar{p}_{\rm out})N$ sites can accept
the car injection.  At the next time step, therefore, the
number of cars injected on the left edge will be
$(1-2\bar{p}_{\rm out})Np$. The equilibrium condition
between the injection and the outflow
\begin{equation}
\bar{p}_{\rm out} = (1-2\bar{p}_{\rm out})p
\end{equation}
gives
\begin{equation}
\bar{p}_{\rm out}={p\over1+2p}\label{lowpout}
\end{equation}
as the average outflow.  This naive estimation of the
outflow (jam-free approximation) agrees well with the
simulation results for $p<p_c$ ($p_c\sim0.2$).  The
extrapolation of eq.~(\ref{lowpout}) to $p=1$ gives
$\bar{p}_{\rm out}=1/3$, which corresponds to the maximum
throughput.

In the high injection $p$ region, the system has traffic jam
clusters in the bulk area.  The outflow $\bar{p}_{\rm out}$
is suppressed and lower than that given by the jam-free
approximation discussed above.

The time dependent behavior of the outflow $p_{\rm out}(t')$
($t'$ denotes the traffic light cycle and $t'=0,\ldots,T-1$)
and its power spectrum
\begin{equation}
I_k=\left\vert\frac{1}{T}\sum_{t'=0}^{T-1}
p_{\rm out}(t')e^{-2\pi i kt'/T}\right\vert.
\end{equation}
is observed, where $T$ is the maximum traffic light cycles
obeying $T=2^\tau< 100N$\cite{FFT}.  In the high $p$ region
(Fig.~\ref{HighDensityPT}), the power spectrum of $p_{\rm
out}(t)$ shows $I_k\sim k^{-\alpha}$ behavior
(Fig.~\ref{P_exponent}).  This shows the existence of the
self-organized jam clusters in the bulk system.  In the low
$p$ region (Fig.~\ref{LowDensityPT}), on the other hand,
$p_{\rm out}(t')$ shows random fluctuation around the
average and the power spectrum is beared with week flicker
noise.

\section{Average velocity}\label{Velocity}
The average velocity of the cars is the number of cars
moving during one traffic light cycle (namely two time
steps).  The arrays $\mu$ and $\nu$ are binary ones.  Thus
the average velocity, naively saying, is half of the Hamming
distance between $\mu(t)$ and $\mu(t+2)$ ($\nu(t)$ and
$\nu(t+2)$).  Caution must be payed to tread the edges of
the system.  For example for right-directed cars, if the
left edge site is empty the site must be excluded from the
calculation of the Hamming distance because of the
probabilistic injection.  If the right edge site is occupied
by right-directed car, the site must be excluded because of
the deterministic outflow.  The same treatment is applied to
up-directed cars.  Figure \ref{AverageVelocity} shows the
results of the simulation.

In the low $p$ region, the jam-free approximation gives the
outflow $\bar{p}_{\rm out}$ as discussed in
\ref{Outflow}. There are $2\bar{p}_{\rm out}N^2$ cars in the
system. By the assumption of no jam in the bulk area, cars
are distributed randomly. There are $\bar{p}_{\rm out}^2N^2$
colliding pairs of cars.  The number of freely moving cars
will be $2\bar{p}_{\rm out}N^2(1-(1/2)\bar{p}_{\rm
out})$. And the average velocity is
\begin{equation}
\bar{v}=1-\frac{1}{2}{p\over1+2p}.
\end{equation}
This estimation agrees with the results of the simulation
less than the case of the outflow.  The discrepancy seems to
come from the effect of collisions with more than two cars.
These effects are expected to decrease faster than two-car
collision in large systems.  The simulation results seems to
show that the jam-free approximation becomes better with
increasing the system size.

At the critical injection $p_c\sim0.2$, the average velocity
shows a sharp phase transition with its sudden decrement by
the formation of jam clusters. It increases gradually with
the increment of $p$ above $p_c$. The behavior of $\bar{v}$
just above $p_c$ shows strong finite size effects.

The time dependent average velocity $v(t')$ and its power
spectrum
\begin{equation}
J_k=\left\vert\frac{1}{T}\sum_{t'=0}^{T-1}
v(t')e^{-2\pi i kt'/T}\right\vert.
\end{equation}
shows the same characteristics as those of the outflow
(Figs.~\ref{LowDensityV} and \ref{HighDensityV}).  The high
$p$ case shows the flicker noise as $J_k\sim k^{-\beta}$
with $\beta\sim1.2$ (Fig.~\ref{V_exponent}) reflecting the
emergence of traffic jam. The low $p$ average velocity seems
to have week flicker noises. This shows the temporary
formations of small traffic jam.

\section{Discussion}\label{Discussion}
In this paper a two-dimensional cellular automaton traffic
flow model with open boundaries was investigated by computer
simulation. The bulk dynamics is deterministic.  Cars are
probabilistically injected from the left and lower sides of
the system and flow out from the right and upper sides
deterministically.

The average outflow $\bar{p}_{\rm out}$, which is the number
of cars flowing out from the right and upper sides per
traffic light cycle and per site, obeys $\bar{p}_{\rm out} =
p/(1+2p)$ in the low injection region ($p<p_c$), where $p$
is the injection rate.  This is well understood with the
jam-free approximation.  High injection $p>p_c$ causes the
emergence of traffic jam clusters in the system and suppress
the outflow.  In the high injection region, the
time-dependent behavior of the outflow shows flicker noises.

The average velocity of cars was also investigated.  The
jam-free approximation value $\bar{v}=1-p/(1+2p)/2$ does not
well agree the simulation results. The reason seems to be
the many car collision effect which will be suppressed in
large system size.  The average velocity shows sharp phase
transition at $p_c\sim0.2$.  In the high injection region,
the average velocity is suppressed by the emergence of
traffic jam clusters.  The time-dependent behavior of the
average velocity also shows flicker noises in the high
injection region.  In the low injection region, it shows
week flicker noise because of the formation of temporal jam
clusters.

In the original BML model, which has periodic boundaries,
cars are freely moving with $\bar{v}=1$ in the low density
region. On the contrary, the current system with open
boundaries shows the average velocity less than unity even
in the jam-free state.  The reason of the difference is as
follows:As well known in one-dimensional case (Wolfram's
rule-184), temporal jam clusters in the low density region
are sorted out and form the maximum throughput current with
the local density $\rho=1/2$.  Sorting out of jam clusters
gives the average velocity $\bar{v}=1$.  In two-dimensional
cases, temporal jam clusters are also sorted out and form
the maximum throughput current with $\rho=2/3$.  In the
periodic boundary cases, once the {\it coherent} maximum
throughput currents are created, they dominated the whole
system and new traffic jam clusters are hardly created. In
the open boundary cases, on the other hand, the {\it
coherent} maximum throughput currents flow out from the
system and the {\it incoherent} currents are injected.
These {\it incoherent} injections form new traffic jam
clusters and suppress the average velocity.

In this paper, I called the event at $p\sim0.2$ as {\it
phase transition}.  The event is not a phase transition in
the strict sense of the word.  No critical behavior is found
at the point.  The value of the average velocity shows sharp
discontinuity at $p\sim0.2$. It has finite value above the
point however.  An adequate order parameter is needed to
strictly define the phase transition.

In Fig.~\ref{AverageVelocity}, the average velocity
increases above the point $p\sim0.4$.  As mentioned above,
the maximum flow with $\rho=2/3$ is formed behind jam
clusters.  The contribution of the maximum flow to the
average velocity is expected to grow with the injection rate
$p$.  The increment of $p$ also seems to enhance the
deterministic feature of the injection process.  These
factors may contribute to the increment of the average
velocity.  On the contrary, the increment of the injection
rate $p$ contributes the formation of jam clusters which
decreases the average velocity.  The competition of this
cluster formation effect and the previous two factors
decides the behavior of the average velocity.  The
investigations of the statistical and dynamical properties
of spatial structures may clarify the behavior of the
system.

Tadaki and Kikuchi shows the existence of two types of jam
phases in the BML model\cite{Tadaki:1,Tadaki:2}.  The
current model with open boundaries seems to have only one
jam phase. The high density random jam phase found in
periodic boundary case seems to be one of the finite size
effects.  In viewpoints of statistical mechanics, finite
size effects will be neglected in realistic macroscopic
systems.  On the contrary, real traffic network systems are
finite and the finiteness may be important factor of the
system.  Observations of real traffic network systems are
expected.

\begin{figure}
\caption{A snapshot of the simulation.  The system size is
$100\times100$ and $p=0.4$.  The black and gray dots show
right-directed and up-directed cars respectively.  There are
some local jam clusters. Jam clusters are sorted out with
the maximum throughput ($\rho=2/3$).  }
\label{snap}
\end{figure}

\begin{figure}
\caption{Average outflow from the system for the system size
$50\times50$, $100\times100$, $200\times200$ and
$400\times400$.  The bold line gives jam-free approximation
$\bar{p}_{\rm out}=p/(1+2p)$ which is given in the text.}
\label{AverageOutflow}
\end{figure}

\begin{figure}
\caption{Time dependent behavior of $p_{\rm out}(t)$ for
$p=0.1$ with the system size $400\times400$ and its power
spectrum $I_k$.  The bold line is fitted with the power
spectrum $10^1<k<10^3$ by the method of least square. 
}
\label{LowDensityPT}
\end{figure}
\begin{figure}
\caption{Time dependent behavior of $p_{\rm out}(t)$ for
$p=0.6$ with the system size $400\times400$ and its power
spectrum $I_k$. The bold line is fitted with the power
spectrum $10^1<k<10^3$ by the method of least square. }
\label{HighDensityPT}
\end{figure}

\begin{figure}
\caption{The behavior of the exponent $\alpha$.  The
exponents are calculated to fit the power spectrum $I_k$
within $10^1<k<10^3$ with the method of lease square.  In
the high $p>p_c$ region, the exponent behaves constant
$\alpha\sim 0.8$ which depends on the system size.  Below
the critical $p_c$ the exponent sharply decrease because the
low $p$ power spectrum shows week flicker noise.}
\label{P_exponent}
\end{figure}

\begin{figure}
\caption{Average velocity for the system size
$50\times50$, $100\times100$, $200\times200$ and $400\times400$.
The bold line is $\bar{v}=1-(1/2)p/(1+2p)$.}
\label{AverageVelocity}
\end{figure}

\begin{figure}
\caption{Time dependent behavior of $v(t)$ for $p=0.1$ with
the system size $400\times400$ and its power spectrum $J_k$. 
The bold line is fitted with the power spectrum
$10^1<k<10^3$ by the method of least square. }
\label{LowDensityV}
\end{figure}
\begin{figure}
\caption{Time dependent behavior of $v(t)$ for $p=0.6$ with
the system size $400\times400$ and its power spectrum $J_k$. 
The bold line is fitted with the power spectrum
$10^1<k<10^3$ by the method of least square. }
\label{HighDensityV}
\end{figure}

\begin{figure}
\caption{The behavior of the exponent $\beta$.  The
exponents are calculated to fit the power spectrum $J_k$
within $10^1<k<10^3$ with the method of lease square.  In
the high $p>p_c$ region, the exponent behaves constant
$\beta\sim 1.2$ which depends on the system size.  Below the
critical $p_c$ the exponent sharply decrease because the low
$p$ power spectrum shows week flicker noise.}
\label{V_exponent}
\end{figure}

\end{document}